\documentclass[sigconf]{acmart}

\AtBeginDocument{%
  \providecommand\BibTeX{{%
    \normalfont B\kern-0.5em{\scshape i\kern-0.25em b}\kern-0.8em\TeX}}}

\copyrightyear{2021} 
\acmYear{2021} 
\setcopyright{acmcopyright}
\acmConference[KDD '21]{Proceedings of the 27th ACM SIGKDD Conference on Knowledge Discovery and Data Mining}{August 14--18, 2021}{Virtual Event, Singapore}
\acmBooktitle{Proceedings of the 27th ACM SIGKDD Conference on Knowledge Discovery and Data Mining (KDD '21), August 14--18, 2021, Virtual Event, Singapore}
\acmPrice{15.00}
\acmDOI{10.1145/3447548.3467101}
\acmISBN{978-1-4503-8332-5/21/08}

\usepackage{makecell}
\usepackage{subfigure}
\usepackage{CJK}
\begin{document}
\fancyhead{}
\begin{CJK*}{UTF8}{gbsn}
\title{Embedding-based Product Retrieval in Taobao Search}

\author{Sen Li, Fuyu Lv}
\authornote{Equal contribution}
\author{Taiwei Jin, Guli Lin, Keping Yang, Xiaoyi Zeng}
\affiliation{%
  \institution{Alibaba Group}
  \city{Hangzhou}
  \country{China}
  \institution{\\ \{lisen.lisen,fuyu.lfy,taiwei.jtw,\\
    guli.lingl,shaoyao,yuanhan\}@alibaba-inc.com}
}

\author{Xiao-Ming Wu$^1$, Qianli Ma$^2$}
\affiliation{%
  \institution{$^1$The Hong Kong Polytechnic University}
  \city{Hong Kong}
  \country{China}
}
\affiliation{
    \institution{$^2$South China University of Technology}
    \city{Guangzhou}
    \country{China}
}
\email{csxmwu@comp.polyu.edu.hk, qianlima@scut.edu.cn}

\renewcommand{\shortauthors}{Li and Lv, et al.}

\begin{abstract}
Nowadays, the product search service of e-commerce platforms has become a vital shopping channel in people's life. The retrieval phase of products determines the search system's quality and gradually attracts researchers' attention. Retrieving the most relevant products from a large-scale corpus while preserving personalized user characteristics remains an open question.
Recent approaches in this domain have mainly focused on embedding-based retrieval (EBR) systems. However, after a long period of practice on Taobao, we find that the performance of the EBR system is dramatically degraded due to its: (1) low relevance with a given query and (2) discrepancy between the training and inference phases. Therefore, we propose a novel and practical embedding-based product retrieval model, named Multi-Grained Deep Semantic Product Retrieval (MGDSPR). Specifically, we first identify the inconsistency between the training and inference stages, and then use the softmax cross-entropy loss as the training objective, which achieves better performance and faster convergence. Two efficient methods are further proposed to improve retrieval relevance, including smoothing noisy training data and generating relevance-improving hard negative samples without requiring extra knowledge and training procedures. 
We evaluate MGDSPR on Taobao Product Search with significant metrics gains observed in offline experiments and online A/B tests. MGDSPR has been successfully deployed to the existing multi-channel retrieval system in Taobao Search. We also introduce the online deployment scheme and share practical lessons of our retrieval system to contribute to the community.
\end{abstract}



\maketitle

\section{Introduction}
\begin{figure}[h]
    \centering
    \includegraphics[scale=0.29]{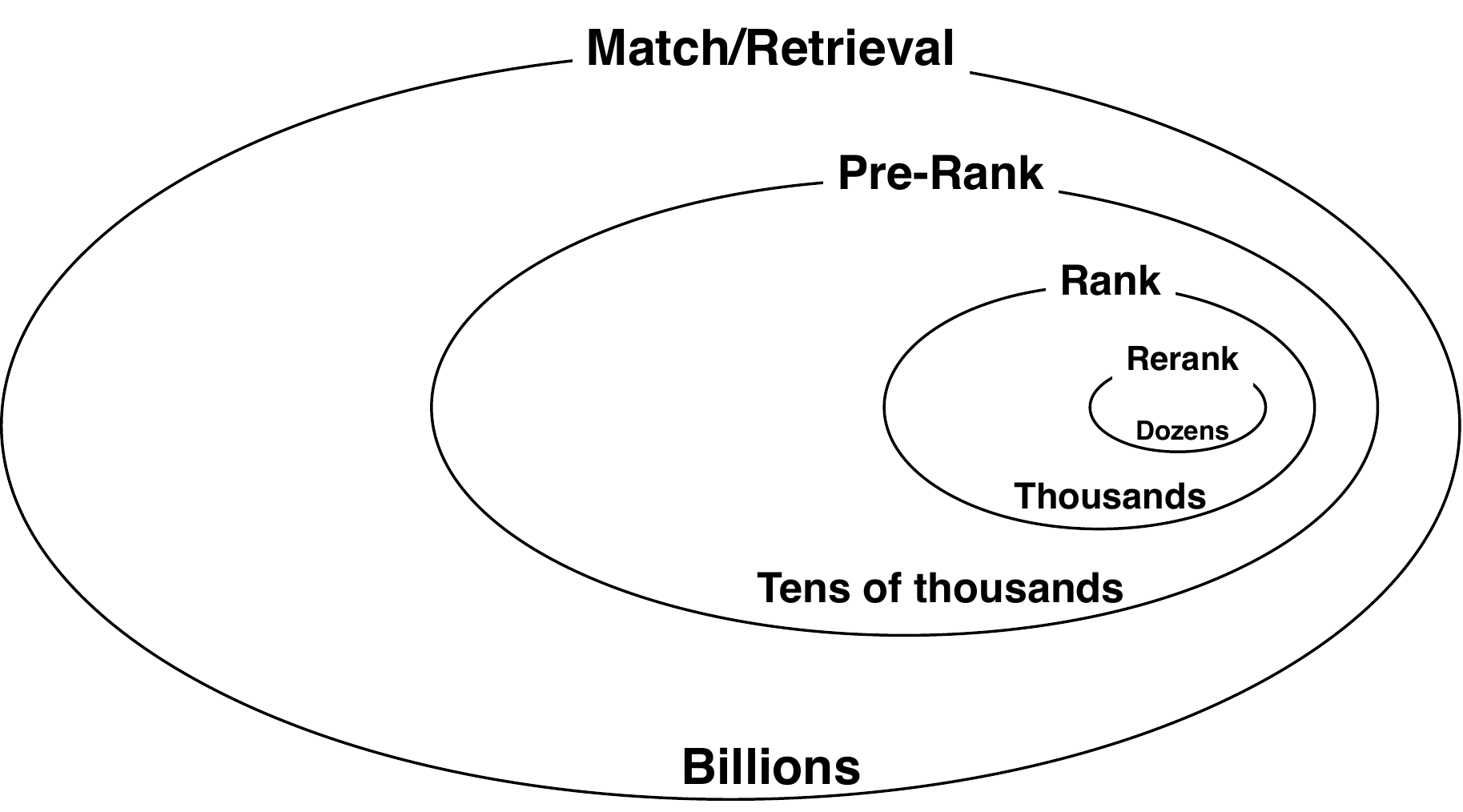}
    \caption{Overview of the product search system in Taobao. The head of each circle denotes different phase. The bottom is the scale of the corresponding candidate set.}
    \label{fig:overview}
\end{figure}
Nowadays, online shopping has become a daily habit in people's lives. 
The top E-commerce shopping platforms (such as eBay, Amazon, Taobao, and JD) have hundreds of millions of daily active users and thus facilitate billion-level transaction records~\cite{sorokina2016amazon,zhang2020towards,liu2017cascade}. 
Therefore, product search engines are designed to discover products that satisfy users, which is also our working goal.
As shown in Figure~\ref{fig:overview}, our search system uses the ``match-prerank-rank-rerank'' architecture to screen and sort thousands of products from billions of candidates to possess controllable and high-efficiency characteristics. We finally return dozens of products to display to users. 
Obviously, the match (retrieval) phase plays an important role in determining the quality of the item candidate set fed to the follow-up ranking stage. 
The problem gradually receives more and more attention from academia and industry.

Search retrieval in e-commerce poses different challenges than in web (document) search: the text in e-commerce is usually shorter and lacks grammatical structure, while it is important to consider the massive historical user behaviors~\cite{ai2017learning,ai2019zero}. 
The lexical matching engine (typically an inverted index~\cite{schutze2008introduction,zobel2006inverted,nigam2019semantic}), despite its widely criticized semantic gap issue~\cite{huang2020embedding,schutze2008introduction,xiao2019weakly}, remains a vital part of current retrieval systems due to its reliability and controllability of search relevance (exactly matching query terms).
However, it hardly distinguishes users' interests in the same query and cannot flexibly capture user-specific characteristics.
Hence, how to effectively retrieve the most relevant products satisfying users while considering the relationship between query semantics and historical user behaviors is the main challenge facing e-commerce platforms.

With the development of deep learning~\cite{zhang2020empowering}, Amazon~\cite{nigam2019semantic} and JD~\cite{zhang2020towards} built their respective two-tower embedding-based retrieval (EBR) systems to provide relevant or personalized product candidates in their e-commerce search engines.
Both of them reported the success of EBR without further discussing its low controllability of search relevance (compared to the lexical matching engine).
We have also built an EBR system that can dynamically capture the relationship between query semantics and personalized user behaviors, and launched it on Taobao\footnote{https://www.taobao.com/} Product Search for quite a long time.
In the first deployment, it can achieve a good improvement in various metrics.
However, after long observation, we have found that the embedding-based method's controllability of relevance is relatively low due to the inability to exactly matching query terms~\cite{guo2016deep}, resulting in increasing user complaints and bad cases that cannot be fixed.
To improve its controllability of relevance (\emph{i.e.}, resolving bad cases), we have adopted a relevance control module to filter the retrieved products.
The control module only keeps those products that meet the relevance standards of exact matching signals and feed them to the follow-up ranking module. However, we statistically find it usually filters out thirty percent of candidates due to the low relevance of retrieved products. It is quite a waste of online computing resources because the filtered products cannot participate in the ranking stage, thus degrading the EBR system's performance.
Therefore, the practical challenge for our search system is to enable the embedding-based model to retrieve more relevant products and increase the number of participants in the subsequent ranking stage.

Moreover, random negative samples are widely used to train large-scale deep retrieval models to ensure the sample space in training is consistent with that of the inference phase~\cite{huang2020embedding,zhang2020towards}. Nevertheless, there remains a discrepancy in existing e-commerce product search methods~\cite{nigam2019semantic,zhang2020towards} due to the inconsistent behavior between the training and inference stages. Specifically, during inference, the model needs to select the top-$K$ products closest to the current query from all candidates, requiring the ability for global comparison. However, \cite{nigam2019semantic} and \cite{zhang2020towards} both adopt hinge (pairwise) loss as the training objective, which can only do local comparison. 

This paper introduces the design of the proposed Multi-Grained Deep Semantic Product Retrieval (MGDSPR) model, its effect on each stage of the search system, and the lessons learned from applying it to product search. To tackle the above problems, we first use the softmax cross-entropy loss as the training objective to equip the model with global comparison ability, making training and inference more consistent. We further propose two effective methods without extra training procedures to enable MGDSPR to retrieve more relevant products. Specifically, we smooth the relevance noise introduced by using user implicit feedback (\emph{i.e.}, click data) logs as training data~\cite{xiao2019weakly,wang2020click} by including a temperature parameter to the softmax function. Also, we mix the positive and random negative samples to generate relevance-improving hard negative samples. Moreover, we adapt the relevance control module to enhance the EBR system's controllability of search relevance.
The effectiveness of MGDSPR is verified by an industrial dataset collected from the Taobao search system and online A/B tests.

The main contributions of this work are summarized as follows:
\begin{itemize}
    \item We propose a Multi-Grained Deep Semantic Product Retrieval (MGDSPR) model to dynamically capture the relationship between user query semantics and his/her personalized behaviors and share its online deployment solution.
    \item We identify the discrepancy between training and inference in existing e-commerce retrieval systems and suggest using the softmax cross-entropy loss as the training objective to achieve better performance and faster convergence.
    \item We propose two methods to make the embedding-based model retrieve more relevant products without additional knowledge and training time. We further adapt the relevance control module to improve the EBR system's controllability of relevance.
    \item Experiments conducted on a large-scale industrial dataset and online Product Search of Taobao demonstrate the effectiveness of MGDSPR. Moreover, we analyze the effect of MGDSPR on each stage of the search system.
\end{itemize}

\section{Related Work}
\subsection{Deep Matching in Search}
With the booming interest in deep NLP techniques, various neural models have been proposed to address the semantic gap problem raised by traditional lexical matching in the last few years.
Those approaches fall into two categories: representation-based learning and interaction-based learning. 
The two-tower structure is the typical characteristic of representation-based models, such as DSSM~\cite{huang2013learning}, CLSM~\cite{shen2014latent},  LSTM-RNN~\cite{palangi2016deep}, and ARC-I~\cite{hu2015convolutional}.
Each tower uses a siamese/distinct neural network to generate semantic representations of query/document. Then a simple matching function (\emph{e.g.}, inner product) is applied to measure the similarity between the query and document.
Interaction-based methods learn the complicated text/relevance patterns between the query and document.
Popular models include MatchPyramid~\cite{pang2016text}, Match-SRNN~\cite{wan2016match}, DRMM~\cite{guo2016deep}, and K-NRM~\cite{xiong2017end}.
Other than semantic and relevance matching, more complex factors/trade-offs, \emph{e.g.}, user personalization~\cite{ai2017learning,ai2019zero,ge2018personalizing} and retrieval efficiency~\cite{covington2016deep}, need to be considered when applying deep models to a large-scale online retrieval system.

\subsection{Deep Retrieval in Industry Search}
Representation-based models with an ANN (approximate near neighbor) algorithm have become the mainstream trend to efficiently deploy neural retrieval models in industry.
For social networks, Facebook developed an EBR system to take text matching and searcher's context into consideration~\cite{huang2020embedding}.
They introduced various tricks and experiences (\emph{e.g.}, hard mining, ensemble embedding, and inverted index-based ANN) to achieve hybrid retrieval (fuzzy matching).
For display advertising, Baidu proposed MOBIUS~\cite{fan2019mobius} for CPM (cost
per mile) maximization in the web ads retrieval phase, reducing the objective distinction between ranking and matching.
For web search, Google~\cite{wu2020zero} 
adopted transfer learning to learn semantic embeddings from data in recommendation systems to alleviate the cold start problem.
Due to more text features and fewer user behaviors, their search scenarios are characterized by strong semantic matching and weak personalization.
For e-commerce search, Amazon developed a two-tower model to address the semantic gap issue in a lexical matching engine for semantic product retrieval~\cite{nigam2019semantic}, where one side uses n-gram query features and the other side exploits item features, without considering user personalization. Recently, JD~\cite{zhang2020towards} proposed a deep personalized and semantic retrieval model (DPSR) to combine text semantics and user behaviors. However, DPSR aggregates user behaviors through average pooling, weakening personalization characteristics.
Furthermore, neither Amazon nor JD studies the problem of insufficient product relevance caused by the EBR method. 
This paper will discuss the low relevance issue of the EBR system encountered in Taobao Product Search and propose our solution.

\begin{figure*}[t] 
    \centering
    \includegraphics[scale=0.27]{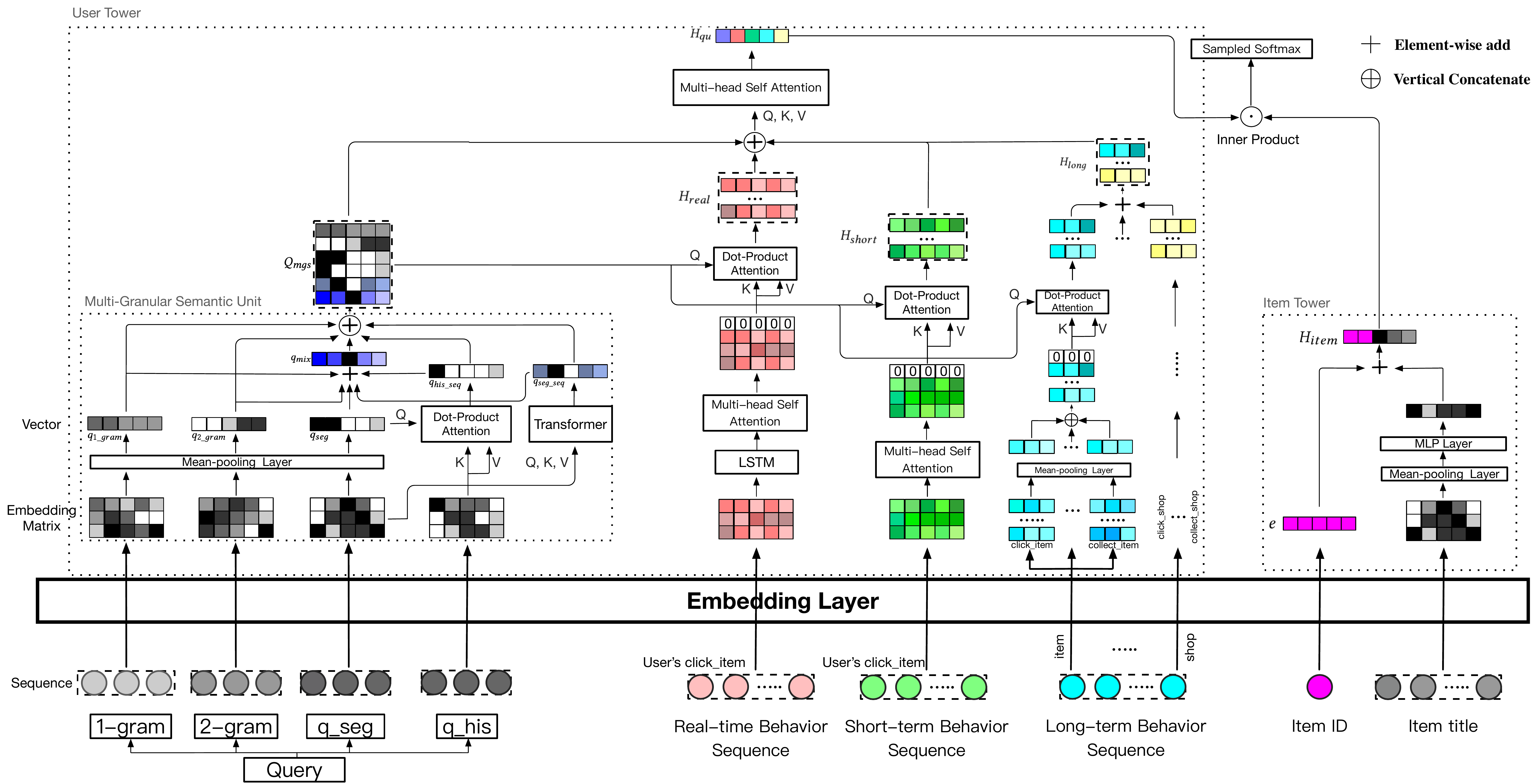}
    \caption{
    General architecture of the proposed
    Multi-Grained Deep Semantic Product Retrieval model (MGDSPR).}
    \label{fig:model}
\end{figure*}

\section{Model}
Here, we introduce our model called Multi-Grained Deep Semantic Product Retrieval (MGDSPR) to simultaneously model query semantics and historical behavior data, aiming at retrieving more products with good relevance. The general structure of MGDSPR is illustrated in Figure~\ref{fig:model}. We first define the problem and then introduce our design of the two-tower model, including the user tower and the item (product) tower. Finally, we elaborate on the training objective and proposed methods to retrieve more relevant products.

\subsection{Problem Formulation}
We first formulate the e-commerce product retrieval problem and our solution as well as the notations used in this paper.
Let $\mathcal{U} = \{u_1, ..., u_u, ..., u_N\} $ denote a collection of $N$ users, $\mathcal{Q} = \{q_1, ..., q_u, ..., q_N\}$ denote the corresponding queries, and $\mathcal{I} = \{i_1, ..., i_i, ..., i_M\}$ denote a collection of $M$ items (products). 
Also, we divide the user $u$'s historical behaviors into three subsets according to the time interval from the current time $t$: real-time (denoted as $\mathcal{R}^u = \{i^u_1, ..., i^u_t, ..., i^u_T\}$, before the current time step), short-term (denoted as $\mathcal{S}^u = \{i^u_1, ..., i^u_t, ..., i^u_T\}$, before $\mathcal{R}$ and within ten days) and long-term sequences (denoted as $\mathcal{L}^u = \{i^u_1, ..., i^u_t, ..., i^u_T\}$, before $\mathcal{S}$ and within one month), where $T$ is the length of the sequence.

We now define the task. Given the historical behaviors ($\mathcal{R}^u, \mathcal{S}^u, \mathcal{L}^u$) of a user $u \in \mathcal{U}$, after he/she submits query $q_u$ at time $t$, we would like to return a set of items $i \in \mathcal{I}$ that satisfy his/her search request. 
Typically, we predict top-$K$ item candidates from $\mathcal{I}$ at time $t$ based on the scores $z$ between the user (query, behaviors) and items, i.e.,
\begin{align}
    z = \mathcal{F}(\phi(q_u, \mathcal{R}^u, \mathcal{S}^u, \mathcal{L}^u), \psi(i)),
\end{align}
where $\mathcal{F}(\cdot)$, $\phi(\cdot)$, $\psi(\cdot)$ denote the scoring function, query and behaviors encoder, and item encoder, respectively. 
Here, we adapt the two-tower retrieval model for efficiency.
We instantiate $\mathcal{F}$ with the inner product function.
In the following, we introduce the design of the user and item towers, respectively.

\subsection{User Tower}
\subsubsection{Multi-Granular Semantic Unit}
Queries in Taobao search are usually in Chinese. After query segmentation, the average length of the segmentation result is less than three. As such, we propose a multi-granular semantic unit to discover the meaning of queries from multiple semantic granularities and enhance the  representation of queries. 
Given a query's segmentation result $q_u = \{w^u_1, ..., w^u_n\}$ (\emph{e.g.}, \{红色, 连衣裙\}), each $w^u = \{c^u_1, ..., c^u_m\}$ (\emph{e.g.}, \{红, 色\}), and its historical query $q_{his} = \{q^u_1, ..., q^u_k\} \in \mathbb{R}^{k \times d}$ (\emph{e.g.}, \{绿色, 半身裙, 黄色, 长裙\}), where $w_n \in \mathbb{R}^{1 \times d}$, $c_m \in \mathbb{R}^{1 \times d}$ and $q_k \in \mathbb{R}^{1 \times d}$, we can obtain its six granular representations $Q_{mgs} \in \mathbb{R}^{6 \times d}$.
It is done by concatenating $q_u$'s unigram mean-pooling $q_{1\_gram} \in \mathbb{R}^{1 \times d}$, 2-gram mean-pooling $q_{2\_gram} \in \mathbb{R}^{1 \times d}$, word segmentation mean-pooling $q_{seg} \in \mathbb{R}^{1 \times d}$, word segmentation sequence $q_{seg\_seq} \in \mathbb{R}^{1 \times d}$, historical query words $q_{his\_seq} \in \mathbb{R}^{1 \times d}$, and mixed $q_{mix} \in \mathbb{R}^{1 \times d}$ representations. 
$d$, $n$, and $m$ denote the embedding size, the number of word segmentation, and the number of words in each segment, respectively.
Formally, the \textbf{M}ulti-\textbf{G}ranular \textbf{S}emantic representation $Q_{mgs}$ is obtained as follows:
\begin{gather}
    q_{1\_gram} =
    mean\_pooling(c_1, ..., c_m),
    \label{eq:q_ugram}
    \\
    q_{2\_gram} =
    mean\_pooling(c_1c_2, ..., c_{m-1}c_m),
    \label{eq:q_tgram}
    \\
    q_{seg} =
    mean\_pooling(w_1, ..., w_n),
    \label{eq:q_seg}
    \\
    q_{seg\_seq} =
    mean\_pooling(Trm(w_1, ..., w_n)),
    \label{eq:seg_seq}
    \\
    q_{his\_seq} = softmax(q_{seg} \cdot (q_{his})^T)q_{his},
    \label{eq:his_seq}
    \\
    q_{mix} = q_{1\_gram}+q_{2\_gram}+q_{seg}+q_{seg\_seq}+q_{his\_seq},
    \label{eq:q_mix}
    \\
    Q_{mgs} = concat(q_{1\_gram},q_{2\_gram},q_{seg},q_{seg\_seq},q_{his\_seq},q_{mix}),
    \label{eq:q_all}
\end{gather}
where $Trm$, $mean\_pooling$, and $concat$ denote the Transformer~\cite{vaswani2017attention}, average, and vertical concatenation operation, respectively. We average all the outputs of the last layer of the Transformer in Eq.~(\ref{eq:seg_seq}).

\subsubsection{User Behaviors Attention}
User behaviors are recorded by their history of items clicked (or bought). Taking user $u$'s short-term behaviors $\mathcal{S}^u$ as an example, $i^u_t \in \mathcal{S}^u$ denotes the user clicks on item $i$ at time $t$, and each item $i$ is described by its ID and side information $\mathcal{F}$ (\emph{e.g.}, leaf category, first-level category, brand and, shop)~\cite{lv2019sdm}. Specifically, each input item $i^u_t \in \mathcal{S}^u$ is defined by:
\begin{gather}
    e^{f}_{i} = W_{f} \cdot {x}_i^{f}
    \label{eq:side_information},
    \\
    i^u_t = concat(\{e^{f}_{i}|f \in \mathcal{F}\}),
    \label{eq:item_hidden}
\end{gather}
where $W_{f}$ is the embedding matrix and ${x_i^{f}}$ is a one-hot vector. $e^{f}_{i} \in \mathbb{R}^{1 \times d_{f}}$ is the corresponding embedding vector of size $d_f$ and $\cdot$ denotes matrix multiplication.
We concatenate the embeddings of item $i$'s ID and side information in Eq.~(\ref{eq:item_hidden}). 
We use the same way to embed items in real-time $\mathcal{R}^u$ and long-term $\mathcal{L}^u$ sequences.

Unlike the target-item attention~\cite{zhou2018deep,zhou2019deep,feng2019deep} used in advertising and recommendation, here we use query attention to capture user history behaviors related to the current query semantics. Moreover, inspired by~\cite{ai2019zero}, we put an all-zero vector into user behavior data to remove potential noise and deal with situations where the historical behaviors may not be related to the current query. In the following, we introduce the fusion of real-time, short-term, and long-term sequences, respectively.

For real-time sequences $\mathcal{R}^u = \{i^u_1, ..., i^u_t, ..., i^u_T\}$, we apply Long Short-Term Memory (LSTM)~\cite{hochreiter1997long} to capture the evolution and collect all hidden states $\mathcal{R}^u_{lstm} = \{h^u_1, ..., h^u_t, ..., h^u_T\}$. Next, we use multi-head self-attention to aggregate multiple potential points of interest~\cite{lv2019sdm} in $\mathcal{R}^u_{lstm}$ to get $\mathcal{R}^u_{self\_att} = \{h^u_1, ..., h^u_t, ..., h^u_T\}$. Then, we add a zero vector at the first position of $\mathcal{R}^u_{self\_att}$, resulting in $\mathcal{R}^u_{zero\_att} = \{0, h^u_1, ..., h^u_t, ..., h^u_T\} \in \mathbb{R}^{(T+1) \times d}$.
Finally, the real-time personalized representation ${H}_{real} \in \mathbb{R}^{6 \times d}$ related to the current query is obtained by the attention operation ($Q_{mgs}$ is analogous to $Q$ in the attention mechanism), which is defined by:
\begin{align}
    H_{real} = softmax(Q_{mgs}\cdot R_{zero\_att}^T)\cdot R_{zero\_att}^T.
\end{align}

For short-term sequences $\mathcal{S}^u = \{i^u_1, ..., i^u_t, ..., i^u_T\}$, we apply multi-head self-attention to aggregate $\mathcal{S}^u$ into $\mathcal{S}^u_{self\_att} = \{h^u_1, ..., h^u_t, ..., h^u_T\}$. We add a zero vector at the first position of $\mathcal{S}^u_{self\_att}$, resulting in $\mathcal{S}^u_{zero\_att} = \{0, h^u_1, ..., h^u_t, ..., h^u_T\} \in \mathbb{R}^{(T+1) \times d}$. Finally, the short-term personalized representation ${H}_{short} \in \mathbb{R}^{6 \times d}$ is defined by:
\begin{align}
    H_{short} = softmax(Q_{mgs}\cdot S_{zero\_att}^T)\cdot S_{zero\_att}^T.
\end{align}
The real-time and short-term sequences are composed of click sequences.

We use four attribute behaviors to describe the long-term sequence (within one month), including \emph{item} ($\mathcal{L}^u_{item}$), \emph{shop} ($\mathcal{L}^u_{shop}$), \emph{leaf category} ($\mathcal{L}^u_{leaf}$) and \emph{brand} ($\mathcal{L}^u_{brand}$). Each attribute behavior is described by a user's click, buy and collecting actions.
For example, $\mathcal{L}^u_{item}$ consists of multiple action sequences: $\mathcal{L}_{click\_item}$, $\mathcal{L}_{buy\_item}$ and $\mathcal{L}_{collect\_item}$. Entries in each action sequence are embedded by Eq.~(\ref{eq:side_information}) and aggregated into a vector through mean-pooling with consideration of quick response in online environment, resulting in $\mathcal{L}^u_{item} = \{0, h_{click},h_{buy},h_{collect}\}$. The  representation of item attribute behavior $H_{a\_item} \in \mathbb{R}^{6 \times d}$ is then defined by:
\begin{align}
    H_{a\_item} = softmax(Q_{mgs}\cdot L_{item}^T)\cdot L_{item}^T.
    \label{eq:u_item}
\end{align}

Finally, the long-term personalized representation ${H}_{long} \in \mathbb{R}^{6 \times d}$ is defined as follows:
\begin{align}
    H_{long} = H_{a\_item} + H_{a\_shop} + H_{a\_leaf} + H_{a\_brand},
    \label{eq:long}
\end{align}
where $H_{a\_shop}$, $H_{a\_leaf}$, and $H_{a\_brand}$ denote the representation of the attribute behaviors of \emph{shop}, \emph{leaf category}, and \emph{brand} respectively.

\subsubsection{Fusion of Semantics and Personalization}
To retrieve products relevant to the current user's query and preserve personalized characteristics, we take the multi-granular semantic representation $Q_{mgs}$ and personalized representations (${H}_{real}, {H}_{short}, {H}_{long}$) as the input of self-attention to dynamically capture the relationship between the two. Specifically, we add a ``[CLS]" token at the first position of the input $I = \{[CLS], Q_{mgs}, {H}_{real}, {H}_{short}, {H}_{long}\}$ of self-attention and regard the output as the user tower's representation $H_{qu} \in \mathbb{R}^{1 \times d}$, which is defined as follows:
\begin{align}
    H_{qu} = Self\_Att^{first}([[CLS], Q_{mgs}, {H}_{real}, {H}_{short}, {H}_{long}]).
    \label{eq:H_qu}
\end{align}

\subsection{Item Tower}
For the item tower, we experimentally use item ID and title to obtain the item representation $H_{item}$. Given the  representation of item $i$'s ID, $e_i \in \mathbb{R}^{1 \times d}$, and its title segmentation result $T_i = \{w^i_1, ..., w^i_N\}$, $H_{item} \in \mathbb{R}^{1 \times d}$ is calculated as follows:
\begin{align}
    H_{item} = e + tanh(W_t\cdot\frac{\sum\nolimits_{i=1}^N w_i}{N}
    ),
\end{align}
where $W_t$ is the transformation matrix. We empirically find that applying LSTM~\cite{hochreiter1997long} or Transformer~\cite{vaswani2017attention} to capture the context of the title is not as effective as simple mean-pooling since the title is stacked by keywords and lacks grammatical structure.

\subsection{Loss Function}
To make the sample space where the model is trained consistent with that of online inference,
~\citeauthor{huang2020embedding}~\cite{huang2020embedding}, ~\citeauthor{nigam2019semantic}~\cite{nigam2019semantic}, and ~\citeauthor{zhang2020towards}~\cite{zhang2020towards}
use random samples as negative samples.
However, they use pairwise (hinge) loss as the training objective, making training and testing behavior inconsistent.
Specifically, during inference, the model needs to pick the top-$K$ items that are closest to the current query from all candidates, which requires the model to have the ability of global comparison. However, hinge loss can only do local comparison. 
Also, hinge loss introduces a cumbersome tuning margin, which has a significant impact on performance~\cite{huang2020embedding}. Here, we adapt the softmax cross-entropy loss as the training objective, achieving faster convergence and better performance without additional hyper-parameter tuning.

Given a user $u$ and his/her query $q_u$, the positive item $i^{+}$ is the item clicked by $u$ under $q_u$. The training objective is defined by:
\begin{gather}
    \hat{y}_{(i^+|q_{u})}= \frac{\exp(\mathcal{F}(q_{u},i^+))}
    {\sum_{i' \in I}\exp(\mathcal{F}(q_{u},i'))},
    \label{eq:softmax}
    \\
    L(\nabla) = -\sum_{i \in I} y_i \log(\hat{y}_i),
\end{gather}
where $\mathcal{F}$, $I$, $i^+$, and $q_u$ denote the inner product, the full item pool, the item tower's representation $H_{item}$, and the user tower's representation $H_{qu}$, respectively.
Note that Eq.~(\ref{eq:softmax}) endows the model with global comparison ability. The softmax involves calculating an expensive partition function, which scales linearly to the number of items. In practice, we use sampled softmax (an unbiased approximation of full-softmax)~\cite{jean2014using,bengio2008adaptive} for training. Similar to~\cite{zhang2020towards}, we also experimentally find that using the same set of random negative samples for every training example in the current batch results in similar performance as using a different set for each one. We adopt the former training method to reduce computing resources.

To improve the EBR system's relevance in retrieval and increase the number of  products participating in the follow-up ranking stage while maintaining high efficiency, we propose two efficient methods without relying on additional knowledge to make our model retrieve more relevant products.

\subsubsection{Smoothing Noisy Training Data}
\label{subsubsec:tau}
In e-commerce search, users' click and purchase records are used as supervisory signals to train a model. However, these signals are noisy since they are influenced not only by query-product relevance but also by images, prices, and user preferences~\cite{xiao2019weakly,wang2020click}.
Hence, we introduce a temperature parameter $\tau$ into softmax to smooth the overall fitted distribution of the training data. If $\tau$->$0$, the fitted distribution is close to one-hot distribution, which means that the model completely fits the supervisory signals. The model will be trained to push positive items far away from negative ones, even if the relevance of a positive item is low. If $\tau$->$\infty$, the fitted distribution is close to a uniform distribution, indicating that the model does not fit the supervisory signals at all. We can increase $\tau$ to reduce the noise in training data and thus alleviate the impact of low relevance caused by fully fitting users' click records, which does not require additional knowledge and is verified by our experiments. Formally, the softmax function with the temperature parameter $\tau$ is defined as follows:
\begin{align}
    \hat{y}_{(i^+|q_{u})}= \frac{\exp(\mathcal{F}(q_{u},i^+)/\tau)}
    {\sum_{i' \in I}\exp(\mathcal{F}(q_{u},i')/\tau)}
    \label{eq:softmax_t}.
\end{align}

\subsubsection{Generating Relevance-improving Hard Negative Samples}
\label{subsec:hard_example}
Unlike prior works ~\cite{nguyen2020learning} that require additional annotated training data and training process, we propose a method to generate relevance-improving hard negative samples in the embedding space.
Specifically, given a training example $(q_u, i^+, i^-)$, where $i^-$ denotes a set of random negative samples sampled from item pool $I$.
For simplicity, we use $q_u$, $i^+$, and $i^-$ to refer to their respective representations. We first select the negative items of $i^-$ that have the top-$N$ inner product scores with $q_u$ to form the hard sample set $I_{hard} $, and then mix $i^+ \in \mathbb{R}^{1 \times d}$ and $I_{hard} \in \mathbb{R}^{N \times d}$ by interpolation to obtain the generated sample set $I_{mix} \in \mathbb{R}^{N \times d}$, which is defined as follows:
\begin{align}
    I_{mix} = \alpha i^+ + (1 - \alpha) I_{hard},
    \label{eq:item_mix}
\end{align}
where $\alpha \in \mathbb{R}^{N \times 1}$ is sampled from the uniform distribution $U(a,b)$ ($0\leq a<b \leq 1$). The closer $\alpha$ is to $1$, the closer the generated sample is to the positive samples $i^+$ in the embedding space, indicating the harder the generated sample is.
We take $I_{mix}$ as the set of relevance-improving hard negative samples and include it in the denominator of the softmax function to make the model distinguish the positive sample $i^+$ and its nearby samples.
Formally, the softmax function with relevance-improving hard samples $I_{mix}$ is defined as follows:
\begin{align}
    \hat{y}_{(i^+|q_{u})}= \frac{\exp(\mathcal{F}(q_{u},i^+)/\tau)}
    {\sum_{i' \in I\cup I_{mix}}\exp(\mathcal{F}(q_{u},i')/\tau)
    }.
    \label{eq:softmax_t}
\end{align}
Note that we can tune the maximum $b$ and minimum $a$ of the uniform distribution $U$ to determine the ``hardness'' of the generated relevance-improving negative samples.
This generation process only needs a linear interpolation after calculating the inner product scores between the current query $q_u$ and the negative samples of $i^-$, which is quite efficient.

\section{System Architecture}
\begin{figure}[h]
    \centering
    \includegraphics[scale=0.37]{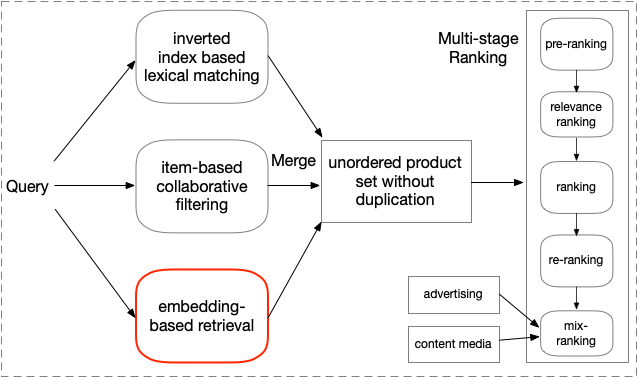}
    \caption{Overview of Taobao search engine.}
    \label{fig:system}
\end{figure}
As illustrated in Figure~\ref{fig:system}, at a high level, the Taobao search engine works as follows: a user issues a query, which triggers a multi-channel retrieval system, producing an unordered candidate set without duplication.
Before the most relevant items are finally displayed to users, the candidates are passed through multi-stages of ranking, including pre-ranking, relevance ranking (removing products that are inconsistent with the predictions of the query's category), ranking, re-ranking, and mix-ranking.
Our embedding-based retrieval module is the third matching channel as a supplement to the existing two-channel retrieval.
In the following, we introduce how to deploy MGDSPR in the production environment once we have trained the model and the relevance control module on the EBR system.

\subsection{Offline Training and Indexing}
\begin{figure}
    \centering
    \includegraphics[scale=0.35]{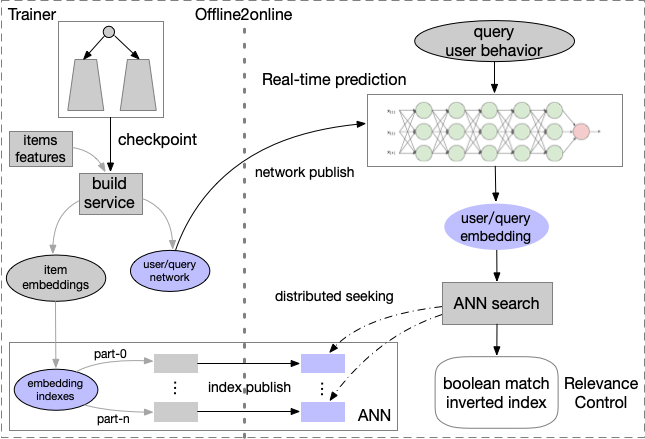}
    \caption{Deployment system of our MGDSPR model.}
    \label{fig:deploy}
\end{figure}

We use search logs in the past one week to train the model by distributed Tensorflow~\cite{abadi2016tensorflow} and update the model parameters daily.
Note that we do not use sequential training~\cite{yi2019sampling} to do the A/B test. Since the base model has been trained by lots of data (several months or even one year), it is difficult to catch up with the same volume of data for the new testing model.
As illustrated in Figure~\ref{fig:deploy}, the deployment system of MGDSPR is an offline to online architecture. 
At the offline phrase, \textit{build service} optimizes and constructs a user/query network extracted from the user tower, which is passed to \textit{real-time prediction} platform. 
All the item embeddings are simultaneously exported from the item tower and transmitted to an approximate near neighbor (ANN) indexing system.
The total number of items is about one hundred millions.
They are placed in multiple columns ($6$ in our system) because of the enormous amounts.
Each column of the ANN builds indexes of embeddings by HC (hierarchical clustering) algorithm with K-means and INT$8$ quantization to promote storage and search efficiency.
The training sample size is $4$ million for HC, and the max scan ratio is $0.01$.

\subsection{Online Serving}
The user/query network and item embedding indexes are published in an online environment after offline indexing.
When a user issues a query, user history behaviors and the query are fed into a real-time prediction platform for online inference. The \textit{ANN search module} then distributively seeks top-$K$ ($K=9600$ in our system) results from indexes of multi-columns (referred to as $n=6$ columns).
Each column returns the same size of $K/n$.
The indexing retrieval accuracy is $98\%$ accompanied with $10$ milliseconds of retrieval latency.

\subsection{Relevance Control}
\label{subsec:rel_cont}
After a long period of practice, we find that although embedding-based retrieval has advantages in personalization and fuzzy matching, it often leads to more search bad cases due to lack of exact matching~\cite{guo2016deeprel} to the key terms of a query.
The \textit{key terms of a query} are referred to as words of brand, type, color, etc., which are significant to product search relevance.
For instance, a user is searching for \textit{Adidas sports shoes}. Items of \textit{Nike sports shoes} are similar to the query in the embedding space and hence will appear in the top-$K$ results with high probability. However, this is not the user intent and will harm user experience.
Hence, we add an inverted index based \textit{boolean matching} module on top of the ANN results. Boolean matching aims to filter out items that do not contain key query terms in their titles. The final search results can then be expressed as: 
\begin{verbatim}
(ANN results) and (Brand: Adidas) and (Category: Shoes).
\end{verbatim}
Generally, we predefine the rule of key terms according to query understanding, \emph{e.g.}, words of brand, color, style, and audience.
Note that Facebook~\cite{huang2020embedding} uses an embedding-based method to enhance boolean expression and achieves fuzzy matching, while we use boolean matching to improve retrieval relevance of the EBR system.
\section{Experiments}
Here, we introduce evaluation metrics, implementation details, datasets, and offline and online experimental results of our method including its effect in the search system.

\subsection{Evaluation Metrics}
\subsubsection{Offline Evaluation}\label{subsubsec:offline_evaluation}
We use the metric of Recall@$K$ to evaluate the offline performance. Specifically, given a query $q_u$, the items clicked or purchased by the user $u$ are regarded as the target set $T = \{t_1, ..., t_N\}$, and the top-$K$ items returned by a model are regarded as the retrieval set $I = \{i_1, ..., i_K\}$. Recall@$K$ is defined as
\begin{align}
    Recall@K = { \sum_{i=1}^{K} i_i \in T \over N}.
    \label{eq:recall}
\end{align}

Empirically, during the retrieval phase, we find that the AUC metric has no positive correlation with the online Gross Merchandise Volume (GMV) metric, while the recall metric does. Also, we add $T_{rec}$, the records relevant to the current query that were not purchased in search but elsewhere (\emph{e.g.}, recommender system) in Taobao Mobile App, to the testing set.

In Taobao search, we also pay attention to the relevance of retrieved products (related to user experience). 
Due to the large amount of test data, we use an online well-trained relevance model (its AUC for human-labeled data is $\textbf{0.915}$) instead of expensive human evaluation to calculate the proportion of products with good relevance (abbreviated as good rate and denoted as $P_{good}$) in the retrieval set $I$,  which is defined as 
\begin{align}
    P_{good}= { \sum_{i=1}^{K} \mathbb I(i_i) \over K},
    \label{eq:good}
\end{align}
where $\mathbb{I}(\cdot)$ is an indicator function. 
When item $i$ is rated as good by the relevance model, the function value is $1$, otherwise $0$.
It is not appropriate to use the AUC metric to evaluate whether our model can return more products with good relevance because it evaluates the order of the set elements rather than the number of the ``good'' elements. $K$ is experimentally set to be $1,000$. 

Meanwhile, to analyze the effect of our model on each stage of the search system, we also count the number of items in the retrieval set $I$ that participate in each follow-up stage.
Given a retrieval set $I = \{i_1, ..., i_{K}\}$, every time it goes through a stage (such as the relevance control module, pre-ranking, and ranking), the number of items will decrease, resulting in $I_{left} = \{i_1, ..., i_{k}\}$, $k < K$. Therefore, we calculate the number of items in $I_{left}$ after going through each phase, and use $Num_{prank}$ and $Num_{rank}$ to denote the number of items that enter the pre-ranking and ranking stages.
For a total retrieval set $\mathcal{I} =\{I_1, ..., I_i, ..., I_N\}$ of $N$ queries, the calculation of $Num_{prank}$ and $Num_{rank}$ are averaged by $N$. 

\subsubsection{Online Evaluation}
We consider the most important online metrics: $\mathbf{GMV}$, $\mathbf{P_{good}}$, and $\mathbf{P_{h\_good}}$.
GMV is the Gross Merchandise Volume, which is defined as
\begin{align}
    GMV = \#\text{pay\,amount}.
    \label{eq:gmv}
\end{align}    

In addition to the amount of online income, we also consider user search experience by the $\mathbf{P_{good}}$ and $\mathbf{P_{h\_good}}$ metrics (defined in Eq.~(\ref{eq:good})). Precisely, both $P_{good}$ and $P_{h\_good}$ calculate the good rate of the item set displayed to users, but $P_{good}$ is determined by the relevance model while $P_{h\_good}$ is determined by humans.

\subsection{Implementation Details}
The maximum length $T$ of real-time, short-term, and long-term sequences are $50$, $100$, and $100$, respectively. We use attention with a mask to calculate those sequences whose length is less than $T$.
The dimensions of the user tower, item tower, behavior sequence, and hidden unit of LSTM are all set to $128$. The batch size is set to $256$.
We use LSTM of two layers with dropout (probability $0.2$) and residual network~\cite{merity2017regularizing} between vertical LSTM stacks. The number of heads in self-attention is set to $8$.
The parameters $a$ and $b$ of uniform distribution $U$ and the number of generated samples $N$ are set to $0.4$, $0.6$ and $684$, respectively. The temperature parameter $\tau$ of softmax is set to $2$. 
All parameters are orthogonally initialized and learned from scratch. The experiments are run on the distributed TensorFlow platform~\cite{abadi2016tensorflow} using $20$ parameter servers and $100$ GPU (Tesla P$100$) workers. The AdaGrad optimizer~\cite{duchi2011adaptive} is employed with an initial learning rate of $0.1$, which can improve the robustness of SGD for training large-scale networks~\cite{dean2012large}. We also adopt gradient clip when the norm of gradient exceeds a threshold of $3$. The training process converges at about $35$ million steps for about $54$ hours.

\subsection{Datasets}
\subsubsection{Large-scale Industrial Offline DataSet}
We collect search logs of user clicks and purchases for $8$ consecutive days from online \textbf{Mobile Taobao App} in December $2020$, and filter the spam users. The training set comprises samples from the first $7$ consecutive days (a total of $4.7$ billion records).
For evaluation, we randomly sample $1$ million search records $T$ and $0.5$ million purchase logs $T_{rec}$ from the recommender system in the $8$-th day. We have also tried to extend the timeframe of training data to $10$ days, but there is no significant benefit, indicating billions of data can effectively prevent the model from overfitting.
The size of the candidate item set is consistent with the online environment, \emph{i.e.}, about $100$ million.

\subsubsection{Online Dataset}
We deploy a well-trained MGDSPR in the Taobao search production environment containing hundreds of millions of user query requests.
The size of the item candidate set is about $100$ million, covering the most active products at Taobao.

\begin{table}[t]
  \centering
  \caption{Comparison with the strong baseline $\alpha$-DNN on a large-scale industrial offline dataset. $Num_{prank}$ is the number of products that flow into the follow-up pre-ranking phase. $P_{good}$ is the good rate. Relative improvements are shown in parentheses.}
  \resizebox{\columnwidth}{!}{
    \begin{tabular}{l|cccc}
    \Xhline{1.2pt}
    Methods & Recall@1000 & $P_{good}$ & $P_{f\_{good}}$ & $Num_{prank}$ \\
    \hline
    $a$-DNN~\cite{covington2016deep}  & 82.6\% & 70.6\% & 83.2\% & 769 \\
    \hline
    MGDSPR & 84.7\%\small{(+2.5\%)} & 80.0\%\small{(+13.3\%)} & 84.1\%\small{(+1.1\%)} & 815\small{(+6.0\%)} \\
    \Xhline{1.2pt}
    \end{tabular}}
  \label{tab:comparison_aDNN}
\end{table}

\subsection{Offline Experimental Results}
Previously, our embedding-based retrieval system adopts the DNN architecture proposed in~\cite{covington2016deep}, but uses more user behaviors and statistical features (inherited from the ranking model), which has been experimentally verified to be effective to some extent. Specifically, we concatenate the vectors of user behaviors (obtained by mean-pooling) and statistical features (\emph{e.g.}, Unique Visitor (UV), Item Page View (IPV)) and feed it into a multi-layer feed-forward neural network. We refer to it as a strong baseline $\alpha$-DNN. In addition, adding statistical features to MGDSPR has no benefit in the metric of $recall$, so we delete them but keep the user behavior sequences.

\subsubsection{Comparison with the Strong Baseline}
As mentioned in Section~\ref{subsubsec:offline_evaluation},
we report the metrics of Recall@$K$, good rate, and $Num_{prank}$. Note that we report $P_{good}$ on both the retrieval set $I$ (denoted as $P_{good}$) and the filtered set $I_{left}$ (denoted as $P_{f\_{good}}$).
As shown in Table~\ref{tab:comparison_aDNN}, 
MGDSPR improves over $\alpha$-DNN by $2.5\%$, $13.3\%$ and $6.0\%$ in Recall@$1000$, $P_{good}$ and $P_{f\_{good}}$ respectively, indicating it can retrieve more products with good relevance and improve the quality of the retrieval set. Comparing $P_{good}$ and $P_{f\_{good}}$ shows our relevance control module enhances retrieval relevance.

\begin{table}[h]
  \centering
  \caption{Ablation study of MGDSPR.}
    \begin{tabular}{l|cc}
    \Xhline{1.2pt}
    Methods & Recall@1000 & $P_{good}$ \\
    \hline
    MGDSPR & 85.6\% & 71.2\% \\
    MGDSPR + \emph{mgs} & 86.0\% & 71.6\% \\
    MGDSPR + \emph{trm} & 86.4\% & 71.4\% \\
    MGDSPR + $\tau$ & 85.5\% & 79.0\% \\
    MGDSPR + \emph{mgs} + \emph{trm} + $\tau$ & 86.8\% & 79.2\% \\
    MGDSPR + $I_{mix}$ & 83.6\% & 75.6\% \\
    \hline
    MGDSPR + all & 84.7\% & 80.0\% \\
    \Xhline{1.2pt}
    \end{tabular}%
  \label{tab:ablation_study}%
\end{table}%

\subsubsection{Ablation Study}
We study the effectiveness of each component of MGDSPR by adding only one component at a time. Specifically, MGDSPR have the following four components: 1) Multi-Granular Semantic unit (\emph{i.e.}, Eq.~(\ref{eq:q_all}), denoted as \emph{mgs}); 2) dynamic fusion of semantics and personalization (\emph{i.e.}, Eq.~(\ref{eq:H_qu}), denoted as \emph{trm}); 3) the temperature parameter $\tau$ of softmax (denoted as $\tau$); 4) the relevance-improving hard negative samples (denoted as $I_{mix}$).
Note that here we focus on the model's performance, so good rate $P_{good}$ is calculated on the retrieval set $I$ instead of $I_{left}$.

As shown in Table~\ref{tab:ablation_study}, both the multi-granular semantics unit \emph{mgs} and \emph{trm} can improve the metrics of Recall@$1000$ and $P_{good}$, indicating the effectiveness of multi-granular semantics and dynamic fusion. The temperature parameter $\tau$ and relevance-improving hard negative samples $I_{mix}$ make the model retrieve more relevant products in terms of much higher good rate $P_{good}$. Comparing MGDSPR+all and MGDSPR or MGDSPR+\emph{mgs}+\emph{trm}+$\tau$, we observe there is a trade-off between recall and relevance even in search scenarios, which may indicate excessive personalization in our system.
\begin{figure}[h]
    \centering
    \includegraphics[scale=0.28]{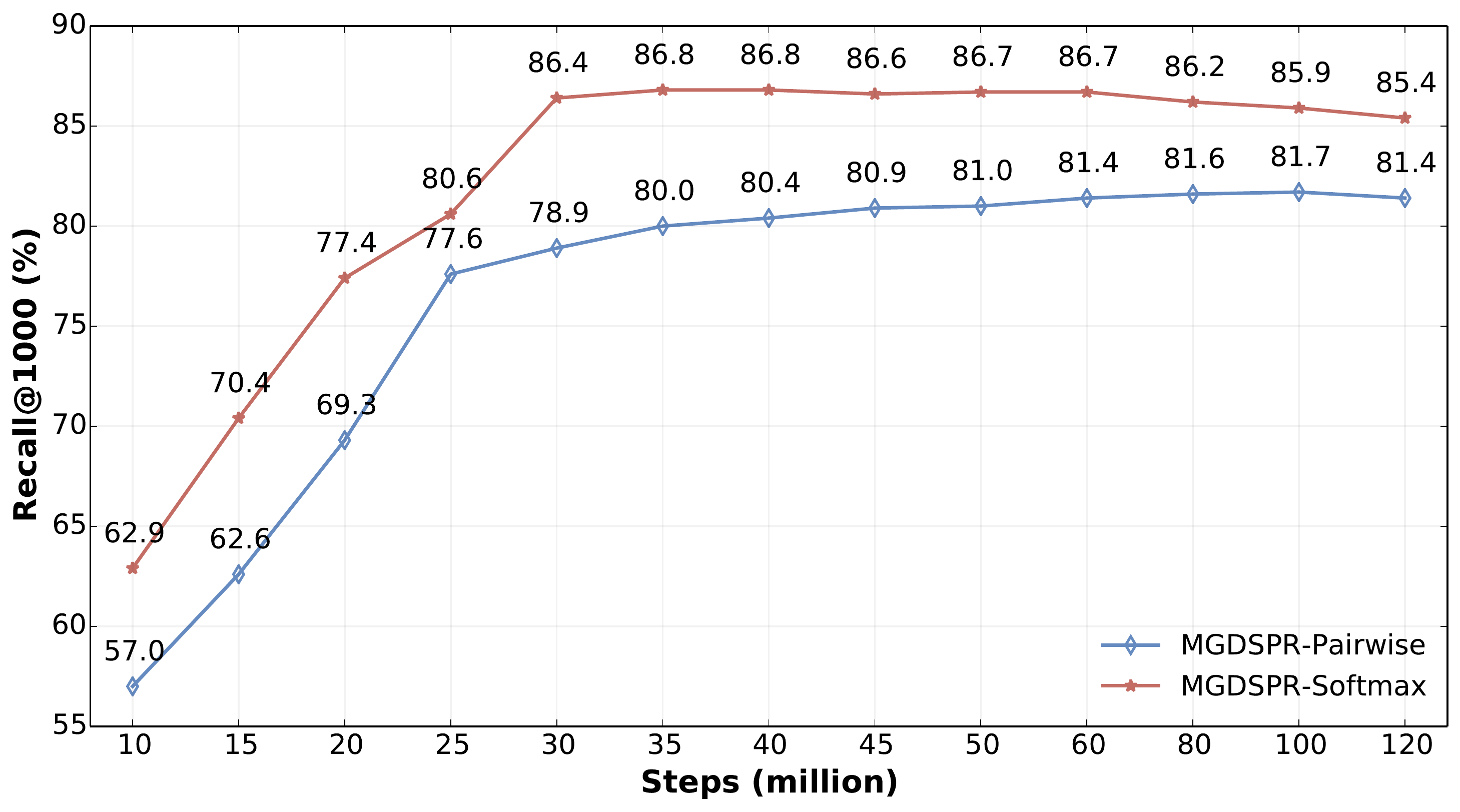}
    \caption{Convergence comparison of the softmax cross-entropy and hinge (pairwise) loss functions. The X-axis denotes the number of training steps, and the Y-axis denotes the corresponding test Recall@$1000$ score.}
    \label{fig:softmax_vs_pairwise}
\end{figure}

\subsubsection{Convergence Analysis}
We investigate the performance of MGDSPR using softmax cross-entropy and pairwise loss~\cite{nigam2019semantic,zhang2020towards} as the training objective, respectively. We report the test Recall@$1000$ score with respect to the nubmer of training steps. As shown in Figure~\ref{fig:softmax_vs_pairwise}, the softmax function's global comparison capability make training and testing more consistent, achieving faster convergence and better performance. In fact, it only takes about three days for the softmax loss to converge while about six days for the pairwise loss. Note that the margin parameter used in the hinge loss has been carefully tuned.

\subsubsection{Hyper-parameter Analysis}
We perform an investigation of the hyper-parameters $\tau$ (for noise smoothing) and $N$ (the number of generated relevance-improving hard negative samples) to demonstrate how they affect the good rate $P_{good}$. We conduct the evaluation by varying $\tau$ (or $N$) while fixing the other parameters.

\begin{figure}[h]
	\subfigure{
        \includegraphics[scale=0.18]{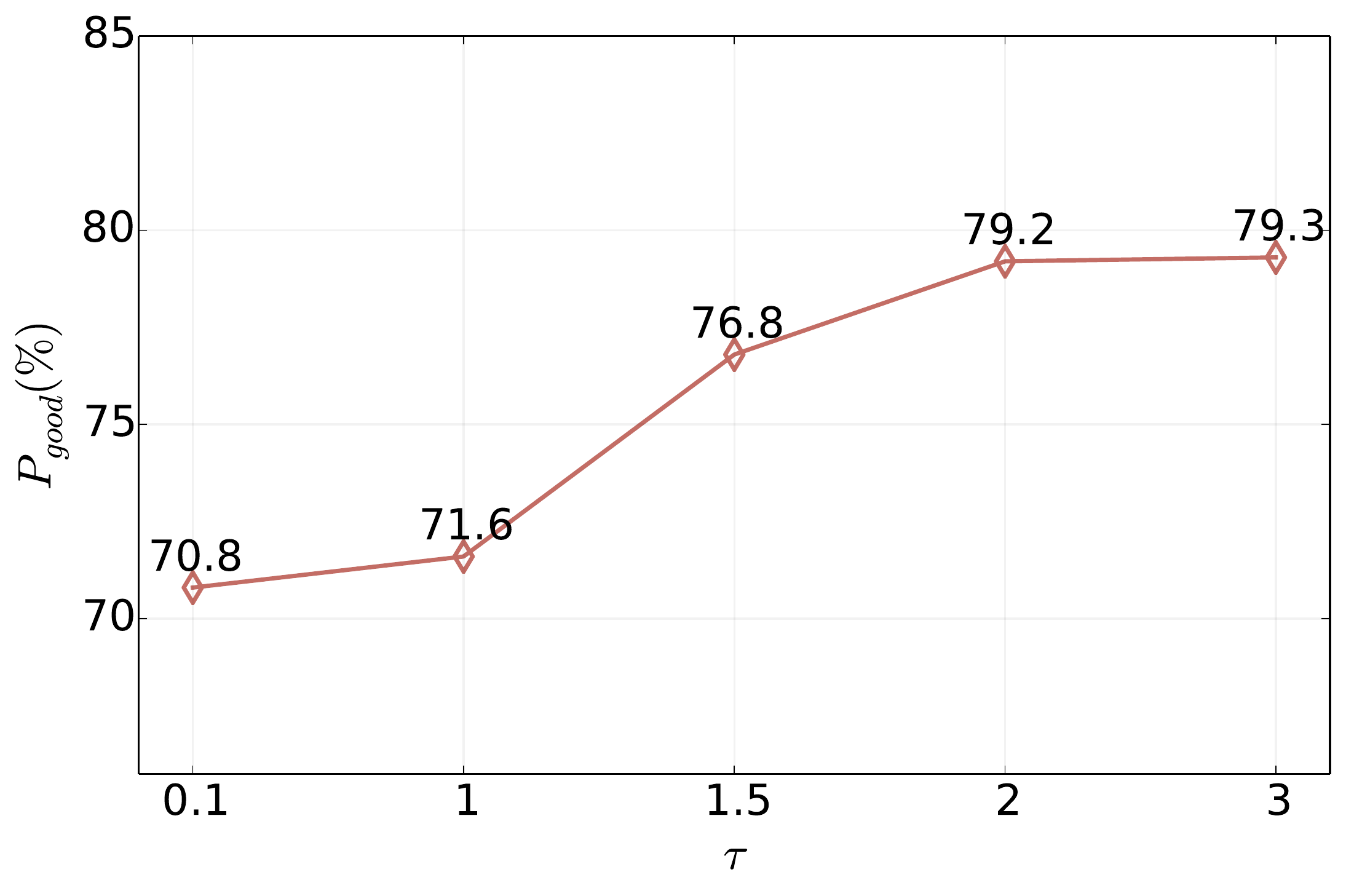}
		}
	\subfigure{
		\includegraphics[scale=0.18]{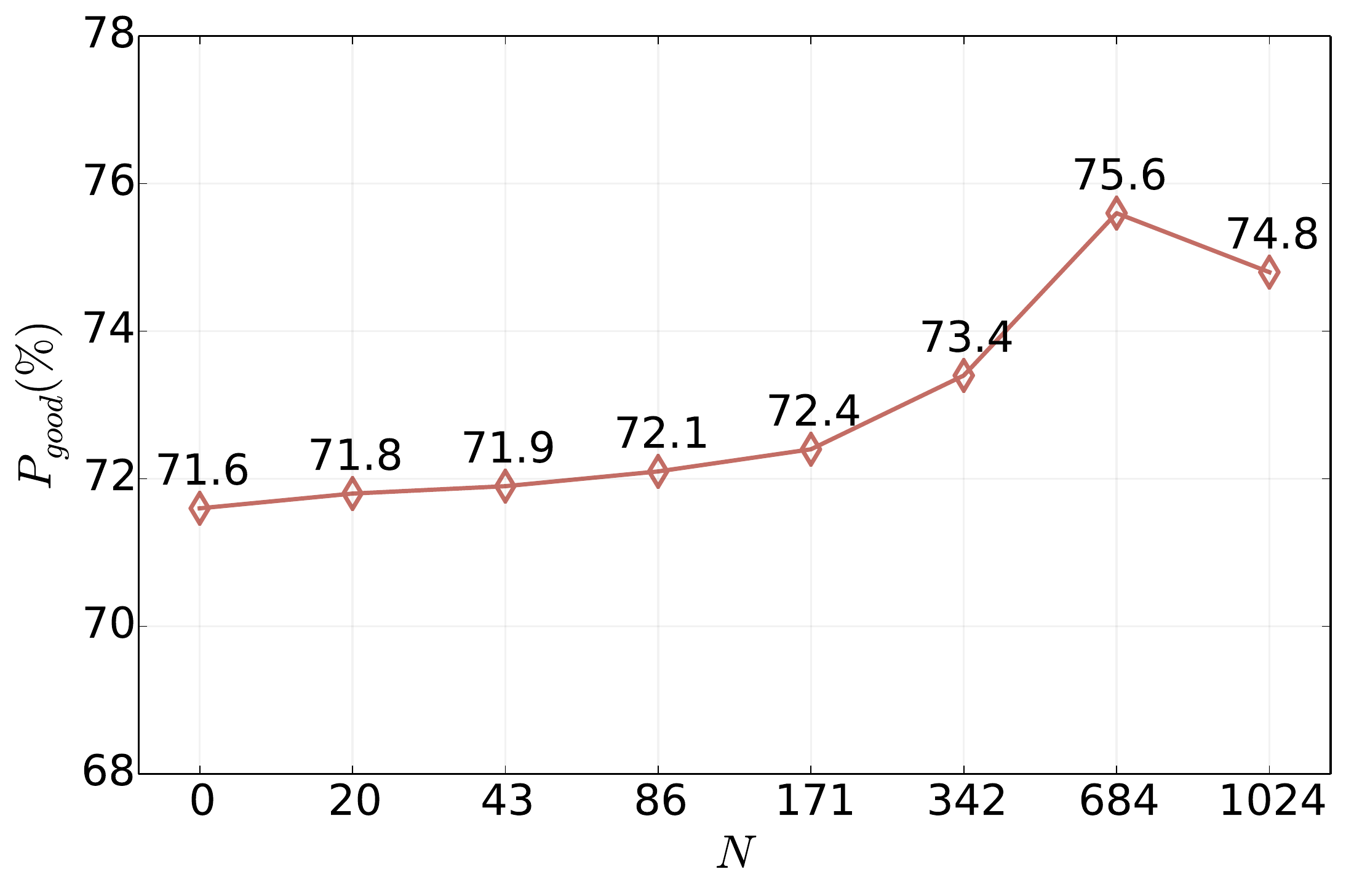}
		}
    \caption{The influence of $\tau$ and $N$ on the good rate $P_{good}$.}
	\label{fig:hyper-parameters}
\end{figure}

As mentioned in Section~\ref{subsubsec:tau}, we can increase $\tau$ to smooth the noisy training data and thus alleviate the effect of insufficient relevance due to overfitting users' click records. As shown in Figure~\ref{fig:hyper-parameters}, $\tau=0.1$ decreases relevance, indicating that the training data does have noise. Also, every non-zero value of $N$ gives better relevance than $N=0$, showing that the generated hard negative samples can improve good rate $P_{good}$. Further, the good rate $P_{good}$ reaches its maximum at $N=684$ and then decreases, indicating that simply increasing the number of samples cannot bring more benefits.

\subsection{Online A/B Test}
We deploy MGDSPR on Taobao Product Search and compare it with the strong baseline $\alpha$-DNN. As aforementioned, to improve user experience, our relevance control module (introduced in Section~\ref{subsec:rel_cont}) will filter out some products retrieved by the EBR system, resulting in low utilization of online computing resources. Therefore, apart from $\mathbf{GMV}$, $\mathbf{P_{good}}$, and $\mathbf{P_{h\_good}}$, we report the number of products that participate in the pre-ranking and ranking phases (denoted as $Num_{prank}$ and $Num_{rank}$) to analyze the model's effect on our search system.
\begin{table}[h]
  \centering
  \caption{The improvements of MGDSPR in $Num_{prank}$, $Num_{rank}$, $P_{good}$, and $P_{h\_good}$ compared with the previous model deployed on Taobao Product Search. The last two columns only report relative values that are calculated on the exposed item set.}
    \begin{tabular}{l|cccc}
    \Xhline{1.2pt}
    Methods & Num$_{prank}$ & Num$_{rank}$ & $P_{good}$ & $P_{h\_{good}}$\\
    \hline
    Baseline & 4070  & 1390  & - & -\\
    \hline
    MGDSPR & 4987\small{(+22.53\%)} & 1901\small{(+36.76\%)} & +1.0\% & +0.35\%\\
    \Xhline{1.2pt}
    \end{tabular}
  \label{tab:online_expriment1}
\end{table}

As shown in Table~\ref{tab:online_expriment1}, after being filtered by the relevance control module, the number of products retrieved by MGDSPR that enter the pre-ranking and ranking phases increases by $22.53\%$ and $36.76\%$, respectively. Obviously, MGDSPR retrieves more products with good relevance and effectively improves the utilization of computing resources. Besides, MGDSPR achieves higher good rates $P_{good}$ and $P_{h\_good}$ of exposure relevance, and thus can display more relevant products to users. 

\begin{table}[h]
  \centering
  \caption{Online A/B test of MGDSPR. The improvements are averaged over $10$ days in Jan $2021$.}
    \begin{tabular}{l|cc}
    \Xhline{1.2pt}
    Launched Platform & GMV   & \#Transactions \\
    \hline
    Taobao Search on Mobile & +0.77\% & +0.33\% \\
    \Xhline{1.2pt}
    \end{tabular}
  \label{tab:online_expriment2}
\end{table}

Finally, we report the $10$-day average of GMV improvements (by removing cheating traffic) achieved by MGDSPR. We also include the corresponding number of transactions (denoted as $\#$Transactions) to increase results confidence. As shown in Table~\ref{tab:online_expriment2}, MGDSPR improves GMV and $\#$Transactions by $0.77\%$ and $0.33\%$, respectively.
Considering the billions of transaction amounts per day in Taobao Search, $0.77\%$ improvement is already tens of millions of transaction amounts, indicating MGDSPR can significantly better satisfy users.

\section{Conclusion}
This paper proposes a practical embedding-based product retrieval model, named Multi-Grained Deep Semantic Product Retrieval (MGDSPR). 
It addresses model performance degradation and online computing resource waste due to the low retrieval relevance in the previous EBR system of Taobao Product Search. 
Meanwhile, we share the lessons learned from solving those problems, including model design and its effect on each stage of the search system, selection of offline metrics and test data, and relevance control of the EBR system.
We verify the effectiveness of MGDSPR experimentally by offline and online A/B tests. 
Furthermore, we have deployed MGDSPR on Taobao Product Search to serve hundreds of millions of users in real time.
Moreover, we also introduce the online architecture of our search system and the deployment scheme of the retrieval model to promote development of the community.

\begin{acks}
We would like to thank the anonymous reviewers for their helpful feedbacks. This research was supported by the Alibaba Innovative Research project $P0034058$ (ZGAL).
\end{acks}

\bibliographystyle{ACM-Reference-Format}
\bibliography{acmart}










\end{CJK*}
\end{document}